\documentclass[10pt,superscriptaddress,twocolumn,showpacs,pre,nofootinbib]{revtex4-1}
\usepackage{bm} 
\usepackage{graphicx} 
\usepackage{amsmath}
\usepackage{amsthm}
\usepackage{amssymb} 
\usepackage{comment} 
\usepackage{epstopdf} 
\usepackage{hyperref}
\usepackage{hypernat}
\usepackage{amsfonts}
\usepackage{color} 
\usepackage{subfigure}
\usepackage[english]{babel}
\usepackage{enumitem}
 \usepackage[bb=boondox]{mathalfa}

\newcommand {\be}{\begin{equation}}
\newcommand {\ee}{\end{equation}}

\newcommand{\ba}{\begin{eqnarray}}
\newcommand{\ea}{\end{eqnarray}}
\newcommand\tr{{\mbox{Tr\,}}}
\newcommand{\ignore}[1]{}

\newcommand{\bes} {\begin{subequations}}
\newcommand{\ees} {\end{subequations}}
\newcommand\D{{\text{c}}}
\newcommand\LT{{\text{L.T.}}}

\DeclareRobustCommand\openzero{\leavevmode{0\kern-.55em0}}
\mathchardef\minus="002D
\usepackage[margin=1in,nofoot]{geometry}

\newcommand{\beq}{\begin{eqnarray}}
\newcommand{\eeq}{\end{eqnarray}}

\newcommand{\e}{{e}}

\newcommand {\bea}{\begin{eqnarray}}
\newcommand {\eea}{\end{eqnarray}}

\newcommand{\bwide}{\begin{widetext}}
\newcommand{\ewide}{\end{widetext}}

\begin{document}

\title{Resolution of the Sign Problem for a Frustrated Triplet of Spins}
\author{Itay Hen}
\affiliation{Information Sciences Institute, University of Southern California, Marina del Rey, California 90292, USA}
\affiliation{Department of Physics and Astronomy and Center for Quantum Information Science \& Technology, University of Southern California, Los Angeles, California 90089, USA}
\email{itayhen@isi.edu}
\begin{abstract}
We propose a mechanism for solving the `negative sign problem'---the inability to assign non-negative weights to quantum Monte Carlo configurations---for a toy model consisting of a frustrated triplet of spin-$1/2$ particles interacting antiferromagnetically. 
The introduced technique is based on the systematic grouping of the weights of the recently developed off-diagonal series expansion of the canonical partition function [Phys. Rev. E 96, 063309 (2017)]. We show that while the examined model is easily diagonalizable, the sign problem it encounters can nonetheless be very pronounced, and we offer a systematic mechanism to resolve it. We discuss the prospects of generalizing the suggested scheme and the steps required to extend it to more general and larger spin models. 
\end{abstract}

\maketitle

\section{Introduction}

The sign problem is the single most important unresolved challenge in quantum many-body simulations.
It appears in a wide variety of areas of physics, chemistry and the material sciences, from superconductivity through neutron stars to lattice quantum chromodynamics and more~\cite{Wiese-PRL-05,marvianHenLidar,signProbSandvik}.
Resolving, or mitigating, the sign problem has therefore rightly been recognized as the holy grail of quantum Monte Carlo techniques since the inception of the field. 

Quantum Monte Carlo (QMC) algorithms~\cite{Alet,PhysRevB.93.054408,PhysRevB.89.134422} are in many cases the only viable method available for studying large quantum many-body systems. The utility of QMC techniques, which evaluate thermal averages of physical observables by the (importance-)sampling of quantum configuration space, hinges on our ability to decompose the partition function of the model of interest into a sum of easily computable non-negative weights, as these in turn are interpreted as probabilities in a Markovian sampling process~\cite{Landau:2005:GMC:1051461,newman}. Whenever terms appear with a negative sign, QMC methods tend to converge exponentially slowly and become essentially impractical. 

In this work we propose a general framework for the possible resolution of the sign problem for spin systems. The approach we take builds on the recently introduced off-diagonal series expansion method~\cite{ODE,ODE2} from which a parameter-free, Trotter error-free series expansion of the partition function of quantum many-body systems is derived. The off-diagonal expansion is carried out around the partition function of the classical component of the Hamiltonian with the expansion parameter being the strength of the off-diagonal component. 

Leveraging the off-diagonal expansion (ODE for short), we consider the resolution of the sign problem for arguably the simplest spin model that possesses it---namely, an antiferromagnetically interacting triplet of spin-$1/2$ particles. As we show, even for this three-spin model the sign problem is strongly manifested, 
prohibiting an efficient evaluation of thermal averages.\footnote{While the toy model in question can be easily diagonalized, we will purposely refrain from resolving the sign problem in this way as a diagonalization approach is not a scalable one.} 
We illustrate how the off-diagonal series expansion can be used towards removing the main hurdle facing quantum Monte Carlo techniques when simulating sign-problematic quantum many-body systems---namely, the existence of negative weights. We then proceed to discuss the prospects of systematically applying the method to larger systems, for which diagonalization is unfeasible. 

We begin by briefly reviewing the off-diagonal partition function series expansion~\cite{ODE,ODE2} followed by an analysis of the emergence of the sign problem and its resolution in the context of the three-spin toy model. We conclude by discussing the generalization of the scheme to large spin systems of physical interest. 

\section{Off-diagonal series expansion}
 
For the sake of brevity, we consider the partition function expansion of  quantum many-body systems whose Hamiltonian can be cast as 
\beq
H = H_\D + \Gamma \sum_j  V_j \,.
\eeq
Here, $H_\D$ is a `classical' Hamiltonian, i.e., a diagonal operator in some known basis, which we refer to as the computational basis, and whose basis states will be denoted by $\{ |z\rangle \}$. 
The $\{V_j \}$ are off-diagonal permutation operators (in the computational basis) that give the system its  `quantum dimension' and obey ${V}_j | z \rangle = | z' \rangle$
for every basis state $|z\rangle$, where $ | z' \rangle \neq |z\rangle$ is also a basis state.\footnote{The partition function expansion can be readily applied to far more general systems. However, for the sake of keeping the derivation short, we shall restrict it to the above simplified version.} The real-valued parameter $\Gamma$ serves as the strength of the quantum component of the Hamiltonian.

The canonical quantum partition function of the above system, $Z=\tr \left[ \e^{-\beta H}\right]$, can be expanded~\cite{ODE,ODE2} in powers of the off-diagonal parameter $\Gamma$ as
\beq
Z  = \sum_{q=0}^{\infty} \Gamma^q\sum_{\{|z\rangle \}} \sum_{\{ {S}_{q}\}} \langle z | {S}_q | z \rangle e^{-\beta[E_{z_0},\ldots,E_{z_q}]} \,.
\label{eq:SSE}
\eeq
Here, $\sum_{\{|z\rangle \}}$ denotes summation over all classical configurations, or basis states $|z\rangle$, and 
\hbox{$\sum_{\{ {S}_{q}\}}$} denotes summation over all distinct products $S_q$ of $q$ off-diagonal operators $V_j$. Each such sequence of operators \hbox{${S}_{q}={V}_{i_1} \cdot {V}_{i_2} \cdots {V}_{i_q}$} is sandwiched between a classical bra $\langle z|$ and a ket $|z\rangle$.
The term $e^{-\beta[E_{z_0},\ldots,E_{z_q}]}$ is the \emph{exponent of divided differences} over the multiset of classical energies $[E_{z_0},\ldots E_{z_q}]$~\cite{dd:67,deboor:05}. 
The energies $E_{z_i}=\langle z_i | H_\D|z_i\rangle$ are the classical energies of the states $|z_0\rangle, \ldots, |z_q\rangle$ obtained from the action of the ordered $V_j$ operators in the sequence ${S}_q$ on $|z_0\rangle$, then on $|z_1\rangle$, and so forth.
Explicitly, $|z_0\rangle=|z\rangle, {V}_{i_1}|z_0\rangle=|z_1\rangle, {V}_{i_2}|z_1\rangle=|z_2\rangle$, etc. 
Since by construction the term $\langle z | {S}_q | z \rangle$ evaluates to either $0$ or to $1$ (the operation $S_q|z\rangle$ returns a basis state $|z'\rangle$ and therefore $\langle z | S_q |z\rangle=\langle z | z'\rangle=\delta_{z,z'}$), the partition function can be more succinctly written as a sum over only non-vanishing terms:
 \beq
Z  =\sum_{\{{S}_{q} : \langle z | {S}_q | z \rangle=1\}}   \Gamma^q e^{-\beta[E_{z_0},\ldots,E_{z_q}]} 
 \,.
\label{eq:SSE2}
\eeq
We interpret the individual terms in the sum above as weights, i.e., $Z  = \sum_{\{\mathcal{C}\}} W_{\mathcal{C}}$, where a configuration $\mathcal{C}$ is a pair $\{ |z\rangle, S_q\}$ whose weight is
\beq \label{eq:gbw}
W_{\mathcal{C}}= \Gamma^q e^{-\beta[E_{z_0},\ldots,E_{z_q}]} \,.
\eeq
We shall refer to $W_{\mathcal{C}}$ as the generalized Boltzmann weight (or GBW) of ${\mathcal{C}}$. 
 
It can be shown~\cite{ODE} that the term $e^{-\beta[E_{z_0},\ldots,E_{z_q}]}$ is positive for even $q$ and negative for odd $q$. 
In order to interpret the $W_{\mathcal{C}}$ terms as actual weights, these must be non-negative~\cite{newman}. The above weights are therefore automatically positive if $\Gamma$ is negative, i.e., if the off-diagonal elements are non-positive, which is the case for `stoquastic' Hamiltonians~\cite{Bravyi:QIC08,Bravyi:2014bf}. As is also evident from the above expression, even values of $q$ yield positive weights regardless of the sign of $\Gamma$ and yield negative values for odd values of $q$ if $\Gamma$ is positive. 

\section{The antiferromagnetic spin triplet} 

\subsection{The model}

\begin{figure}[htp]
\includegraphics[width=0.25\textwidth]{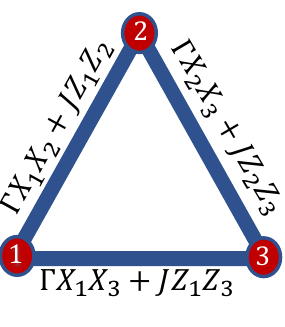}
\small
\caption{\label{fig:triangle} A triplet of antiferromagnetically coupled spin-$1/2$ particles. The two-body interactions are antiferromagnetic in the x-direction with coupling strength $\Gamma$ and in the z-direction with strength $J$. } 
\end{figure}

Having reviewed the ODE partition function expansion, we are now in a position to examine, in that context, the emergence of the sign problem in spin systems. 
 We consider a simple toy model consisting of three antiferromagnetically coupled spin-$1/2$ particles, whose Hamiltonian is given by
\bea\label{eq:H}
H&=&J \left(Z_1 Z_2+Z_2 Z_3+Z_3 Z_1 \right)\\\nonumber
&+& 
\Gamma \left(X_1 X_2+X_2 X_3+X_3 X_1 \right) \,.
\eea
Here, $Z_i$ and $X_i$ for $i=1,2,3$ are the Pauli-z and Pauli-x operators, respectively, acting on the $i$-th spin. The classical part of the Hamiltonian is \hbox{$H_\D=J \left(Z_1 Z_2+Z_2 Z_3+Z_3 Z_1 \right)$} with $J>0$, and the three off-diagonal operators are $V_1=X_2 X_3$, $V_2=X_3 X_1$ and $V_3=X_1 X_2$. A positive value of the off-diagonal parameter $\Gamma$, implying antiferromagnetic coupling along the x-direction, leads to a sign problem, which as we shall illustrate, can be a severe one. The model is depicted in Fig.~\ref{fig:triangle}.

For such a small system, the Hamiltonian is easily diagonalizable and may be readily represented in that basis for arbitrary values of $\Gamma$ and $J$. In this basis there is obviously no sign problem. For the purposes of this study we shall refrain from `curing' the sign problem by a change of basis, as such a strategy is not expected to be feasible for larger systems~\cite{marvianHenLidar,signProbSandvik}. 

\subsection{Emergence of the sign problem}

The computational basis of the three-spin toy model consists of eight basis states. The spectrum of $H_\D$  has two energy levels. The excited states are the two fully aligned configurations {\bf 0}$\equiv|000\rangle$ and {\bf 7}$\equiv|111\rangle$ which have an  energy $E_1=3 J$ (the boldfaced notations {\bf 0} and {\bf 7} are the decimal values corresponding to the binary representations of the two states). On  the other hand, the ground state is six-fold degenerate with $E_0=-J$.  This information is summarized in Table~\ref{tab1}. 
\begin{table*}
\begin{tabular}{|c|c|c|}
\hline
Classical energy $\langle z | H_c | z\rangle$ &States with even parity & States with odd parity   \\
\hline
$E_0=-J$ (ground) & {\bf 3}$\equiv|011\rangle$, {\bf 5}$\equiv|101\rangle$, {\bf 6}$\equiv|110\rangle$ & {\bf 1}$\equiv|001\rangle$,{\bf 2}$\equiv|010\rangle$,{\bf 4}$\equiv|100\rangle$ \\
$E_1=3J$ (excited) & {\bf 0}$\equiv|000\rangle$ & {\bf 7}$\equiv|111\rangle$ \\
\hline
\end{tabular}
\caption{\label{tab1}The eight computational basis states of the antiferromagnetic triplet and their classical energies. The states are divided to two groups of distinct parities.}
\end{table*}

\begin{figure}[htp]
\includegraphics[width=0.35\textwidth]{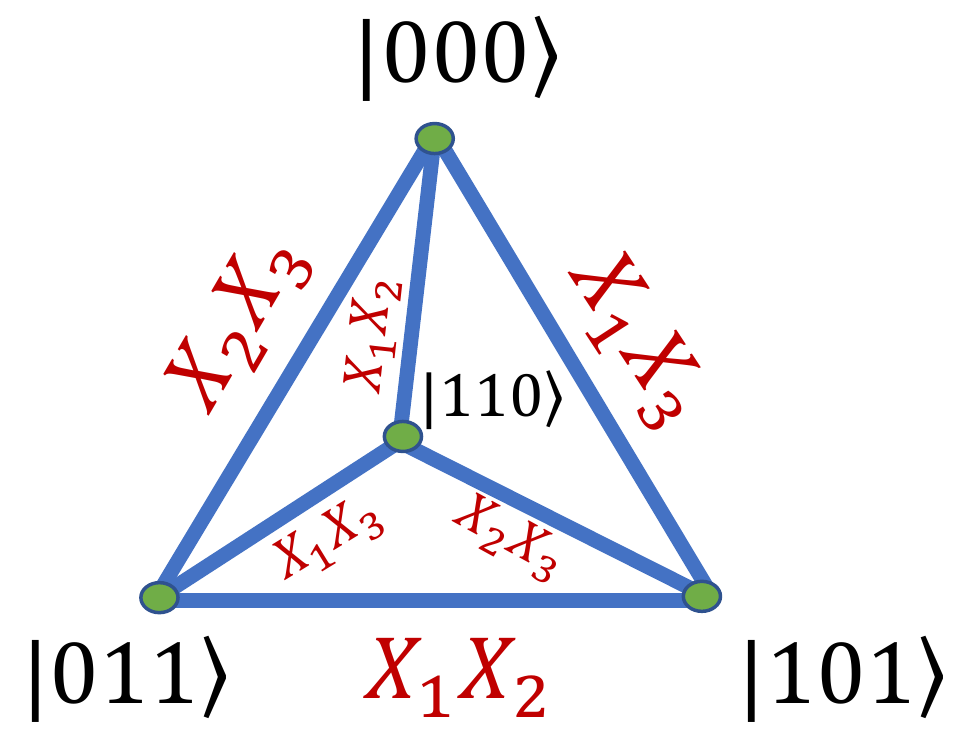}
\small
\caption{\label{fig:tetra} The action of the off-diagonal operators on the even-parity basis states of the spin triplet.  Similar relations hold for the odd-parity states; these are obtained via the substitution $|0\rangle \leftrightarrow |1\rangle$. } 
\end{figure}

\begin{figure*}[htp]
\includegraphics[width=1.75\columnwidth]{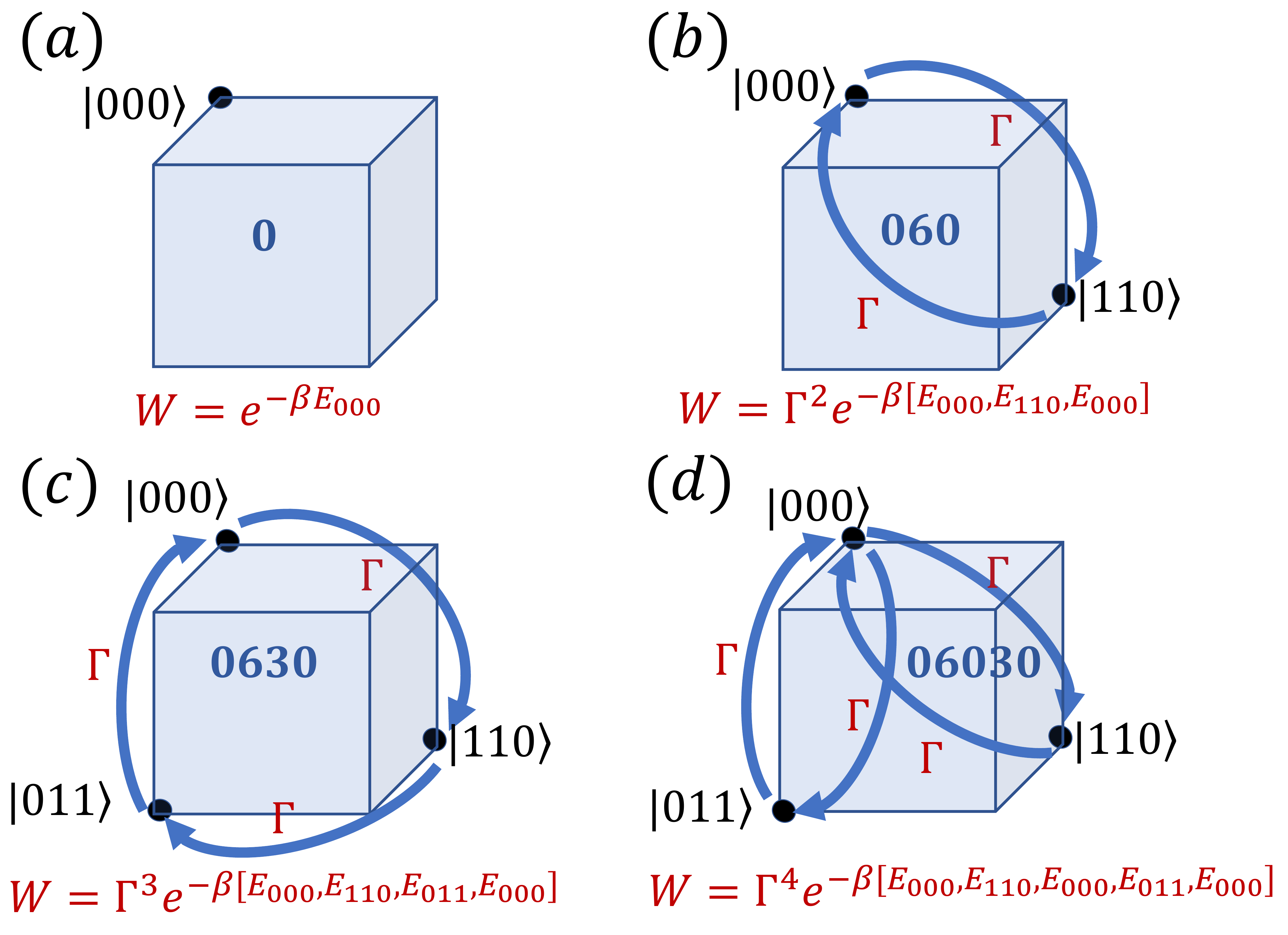}
\caption{\label{fig:qmcWeights} Diagrammatic representation of weights in the partition function expansion as closed paths on the hypercube of basis states. Each configuration is represented by a closed cycle whose nodes are basis states and is assigned a Boltzmann-like weight of the form $\Gamma^q\e^{-\beta[E_{z_0},\ldots,E_{z_q}]}$ calculated from the classical energies $E_{z_i}$ of the classical states $|z_i\rangle$. The boldfaced sequence of digits appearing next to each path corresponds to the sequence of visited basis states. (a) A zeroth-order classical term. These terms appear in the decomposition of the classical partition function. Their weights are standard Boltzmann weights. (b) A second-order term containing three basis states and two edges. (c) A third-order term generating a negative weight (for positive $\Gamma$). (d) A fourth-order term, whose weight is positive regardless of the sign of $\Gamma$.}
\end{figure*}

As noted earlier, an ODE configuration $\mathcal{C}$ consists of a basis state $|z\rangle$ and a product \hbox{${S}_{q}={V}_{i_1} \cdot {V}_{i_2} \cdots {V}_{i_q}$} of off-diagonal operators which together induce a sequence of classical states $|z_i\rangle$ generated by the action of the off-diagonal operators on $|z\rangle$. The sequence of basis states $\{|z_i\rangle \}$ may be viewed as a `path' in the hypercube of basis states (see Fig.~\ref{fig:qmcWeights}). For a weight to have a nonzero value, the path must be a closed one, namely, $|z\rangle=|z_0\rangle=|z_q\rangle$. The actions of the off-diagonal operators $V_1, V_2$ and $V_3$ on the eight basis states of this model are illustrated in Fig.~\ref{fig:tetra}. The off-diagonal operators conserve parity (evenness of number of spins with a given orientation) and therefore only connect  states within a parity sector. Since a configuration can be represented as a closed path on the hypercube with no ambiguity, it will be useful to denote configurations as sequences of the digits {\bf 0}$\ldots${\bf 7}, with each digit signifying a spin configuration along the path (see Table~\ref{tab1}). Examples for ODE configurations and their representations as digit sequences are given in Table~\ref{tab:paths}.

\begin{table}
\begin{tabular}{|c|c|}
\hline
ODE configuration $\mathcal{C}=\{|z\rangle, S_q\}$ & path representation \\
\hline
 $\{|011\rangle, \mathbb{1}\}$ & {\bf 3}\\
 $\{|101\rangle,  V_2 V_2 \}$ &{\bf 505} \\
  $\{|111\rangle,  V_3 V_2 V_1 \}$ &{\bf 7147} \\
$\{|001\rangle,  V_3 V_3 V_2 V_2 \}$& {\bf 17141} \\
\hline
\end{tabular}
\caption{\label{tab:paths}ODE configurations and their representations as closed paths on the hypercube of basis states.}
\end{table}

As discussed above, the classical energies of the basis states $\{|z_i\rangle\}$, $E_{z_i}$, determine the weight of the configuration. The antiferromagnetic triplet has only two energy levels $E_0=-J$ and \hbox{$E_1=E_0+\Delta=3J$} (that is, the classical gap is \hbox{$\Delta=4J$}). In this case, the GBW Eq.~(\ref{eq:gbw}), can be computed analytically. 
A configuration inducing $m_j$ states all with energy $E_j$ (where $j \in \{0,1\}$) yields the GBW:
\beq\label{eq:Wm}
W_{\{E_j^{\otimes m_j}\}}= \frac{(-\beta \Gamma)^{q}}{q!} \e^{-\beta E_j} = \frac{(-\beta \Gamma)^{m_j-1}}{(m_j-1)!} \e^{-\beta E_j}\,. \nonumber\\
\eeq
The weight of a configuration with $m_0>0$ states of energy $E_0$ and $m_1>0$ states with $E_1$ can be similarly calculated to give: 
\begin{widetext}
\beq\label{eq:Wm0m1}
W_{\{E_0^{\otimes m_0},E_1^{\otimes m_1}\}}=\Gamma^q \e^{-\beta E_0} \frac{(m_0+m_1)!\left[{}_1F_1(1-m_0,1-q,-\beta \Delta)-\e^{-\beta \Delta} {}_1F_1(1-m_1,1-q,\beta \Delta )\right]}{(-1)^{m_1}\Delta^q(m_0-1)!(m_1-1)!} \,,
\eeq
\end{widetext}
where $q=m_0+m_1-1$ is the number of off-diagonal operators in the sequence $S_q$ and $ {}_1F_1$ is the Kummer confluent hypergeometric function (a detailed derivation of the two equations above is given in App.~\ref{app:wc}).
The weights Eqs.~(\ref{eq:Wm}) and~(\ref{eq:Wm0m1}) are indeed negative for any $\Gamma>0$ and odd $q$, indicating the emergence of a sign problem for this model. 

For reasons that will become clear later, we calculate the number of distinct configurations (of a given parity) with energy multiplicities $(m_0,m_1)$. It is given by
\bea
&N&_{(m_0,m_1)}=- 2(-1)^{m_0} \delta_{0,m_1}
+ \frac{3^{m_1} 2^{m_0-m_1}(m_0-1)!}{6 (m_0-m_1+1)! m_1!}\nonumber\\
&\times& \left[
4 m_1(m_1-1)+3(m_0-m_1+1)(m_0-m_1)  \right]\,.
\eea
A full derivation of the above expression is given in App.~\ref{app:Nm0m1}. 
Table~\ref{tab:Nm0m1} provides the explicit count for the first few $(m_0,m_1)$ sectors alongside some sample configurations. 
\begin{table}
\begin{tabular}{|c|c|c|}
\hline
multiplicities & number of  config-& examples \\
$(m_0,m_1)$ & urations $N_{(m_0,m_1)}$& (even parity sector) \\
\hline
\hline
(1,0) & 3 & {\bf 3}, {\bf 5}, {\bf 6}. \\
(2,0) & 0 & --- \\
(3,0) & 6 & {\bf 353}, {\bf 565}, $\ldots$ \\
(4,0) & 6 & {\bf 3563}, {\bf 6356}, $\ldots$ \\
(5,0) & 18 & {\bf 35653}, {\bf 56565}, $\ldots$ \\
$\ldots$ & $\ldots$ & $\ldots$ \\
\hline
(0,1) & 1 & {\bf 0} \\
(1,1) & 0 & --- \\
(2,1) & 3 & {\bf 303}, {\bf 505}, {\bf 606}.\\
(3,1) & 12 & {\bf 3503}, {\bf 5065}, $\ldots$\\
(4,1) & 36 & {\bf 36503}, {\bf 53065}, $\ldots$\\
$\ldots$ & $\ldots$ & $\ldots$ \\
\hline
(1,2) & 3 & {\bf 030}, {\bf 050}, {\bf 060}.\\
(2,2) & 6 & {\bf 0350}, {\bf 0650}, $\ldots$ \\
(3,2) & 21 & {\bf 03530}, {\bf 06530}, $\ldots$ \\
(4,2) & 78 & {\bf 035650}, {\bf 065360}, $\ldots$ \\
$\ldots$ & $\ldots$ & $\ldots$ \\
\hline
(2,3) & 9 & {\bf 03050}, {\bf 03060}, $\ldots$ \\
$\ldots$ & $\ldots$ & $\ldots$ \\
\hline
\end{tabular}
\caption{\label{tab:Nm0m1}Number of distinct configurations (within a parity sector) for various energy multiplicities $(m_0,m_1)$.  Only even-parity configurations are given as examples. For any fixed $m_1$, $N_{(m_0,m_1)}$ 
grows asymptotically exponentially with $m_0$. }
\end{table}

To measure the severity of the sign problem in QMC, it is useful to study the quantity
\beq
\langle \text{sgn}\rangle=\frac{\sum_{\mathcal{C}} W_{\mathcal{C}}}{\sum_{\mathcal{C}} |W|_{\mathcal{C}}}\,,
\eeq 
which may also be written as:
\beq
\langle \text{sgn}\rangle= \frac{\sum_{\mathcal{C}} \text{sgn}(W) |W|_{\mathcal{C}}}{\sum_{\mathcal{C}} |W|_{\mathcal{C}}}= \langle \text{sgn}(W)\rangle_{|W|}\,.
\eeq 
Here, $\langle \text{sgn}(W)\rangle_{|W|}$ denotes the Monte Carlo thermal average with respect to the absolute values of the weights $W_{\mathcal{C}}$. The quantity $\langle \text{sgn}\rangle$ may therefore be viewed as the thermal average of the sign of the weight (with respect to the distribution of absolute weights), or the `weighted sign' for short.  The weighted sign also appears in the evaluation of thermal averages of physical observables via the relation. 
\beq\label{eq:A}
\langle A\rangle = \frac{\langle A \, \text{sgn}(W)\rangle_{|W|}}{\langle \text{sgn}(W)\rangle_{|W|}} = \frac{\langle A \, \text{sgn}(W)\rangle_{|W|}}{\langle \text{sgn} \rangle} \,,
\eeq 
meaning that the thermal average of a physical observable $A$ is a ratio of two quantities that are thermal averages with respect to the distribution of absolute weight.\footnote{It is interesting to note that the absolute weights are precisely those obtained for the sign-problem-free model with $\Gamma \to -\Gamma$.}

\begin{figure}[htp]
\includegraphics[width=0.45\textwidth]{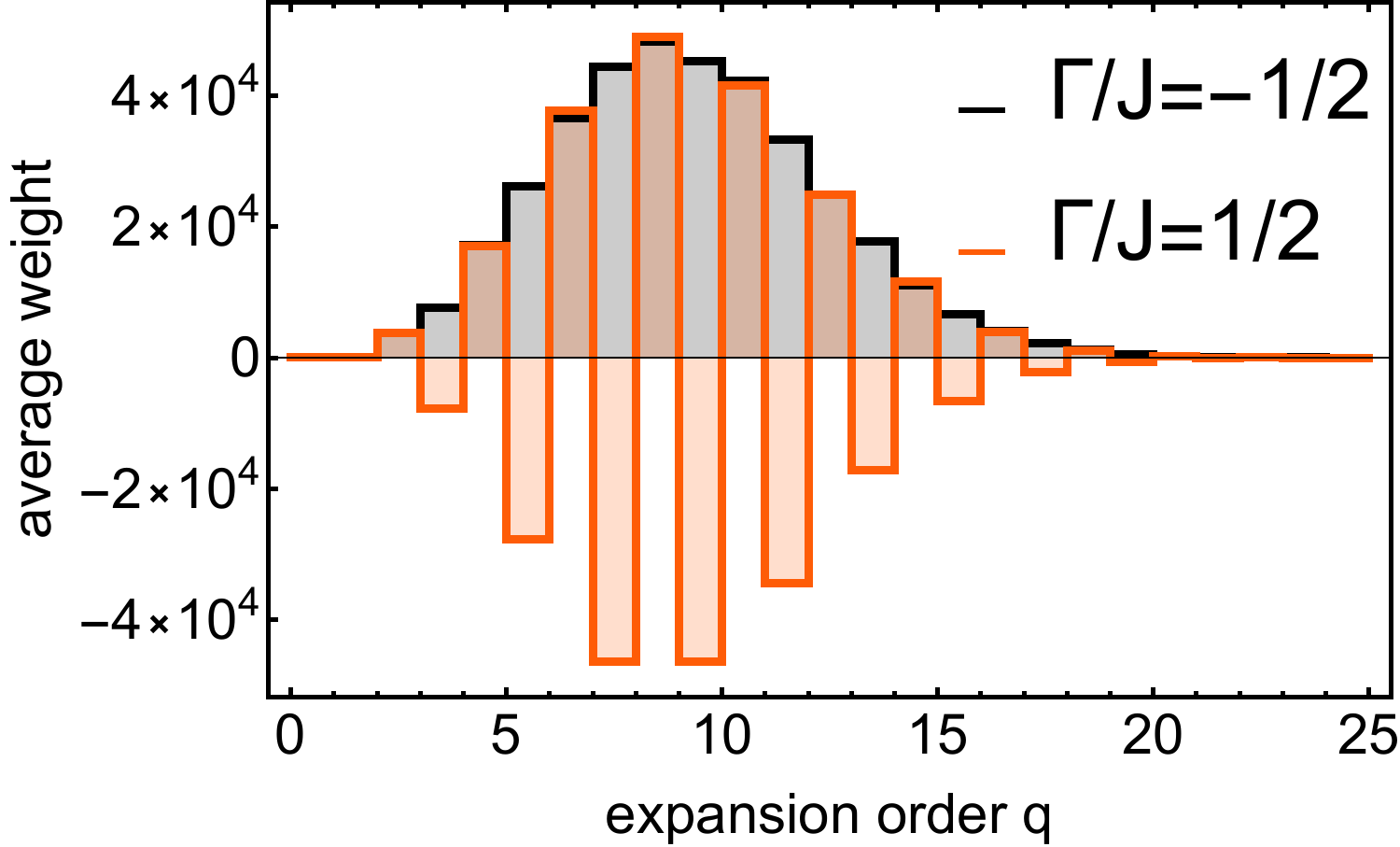}
\includegraphics[width=0.45\textwidth]{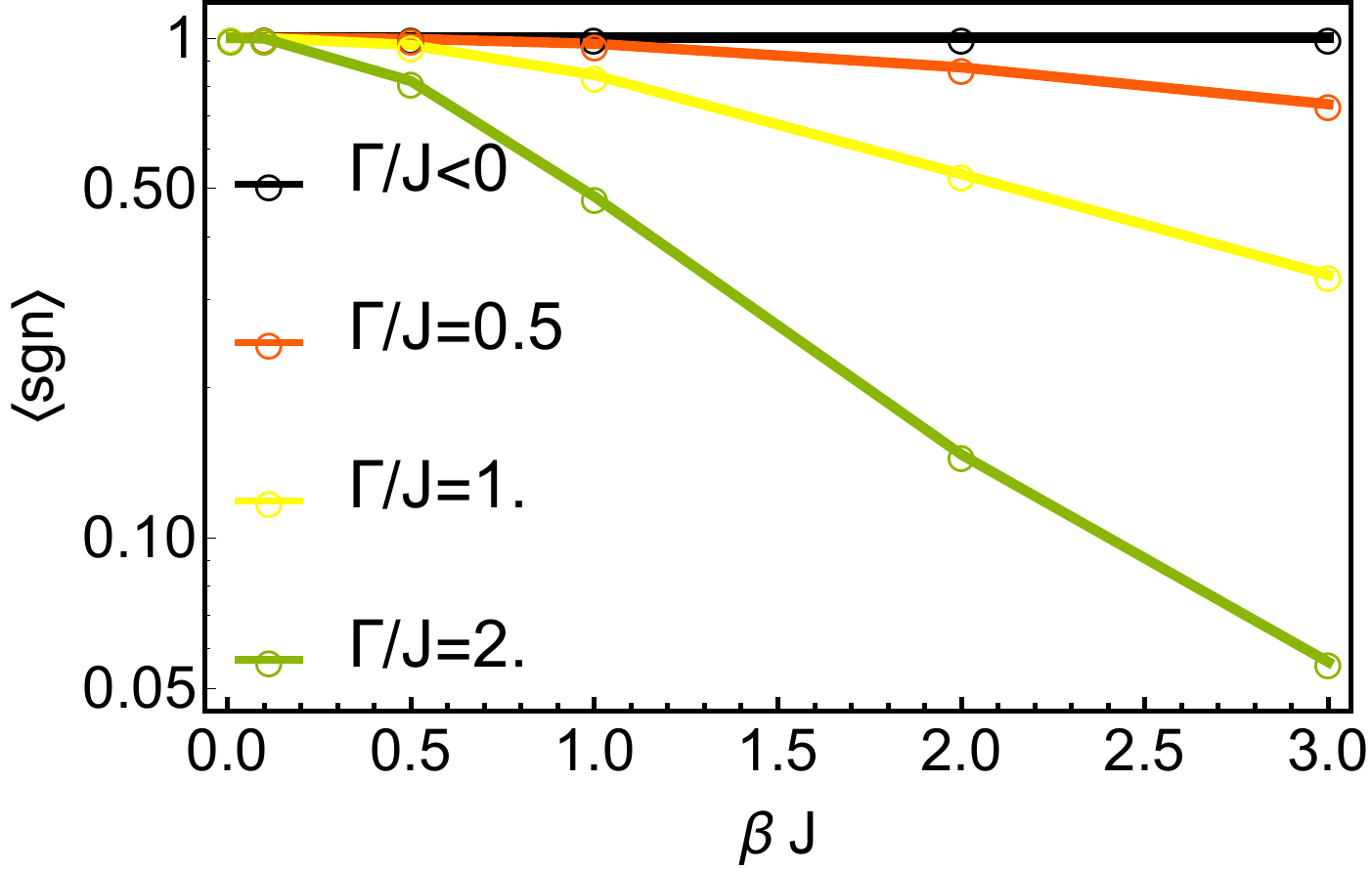}
\small
\caption{\label{fig:sign} Top: Average QMC weight as a function of expansion order $q$ in the sign-problem-free case (black) and the sign problematic case (red). The latter distribution has negative weights for odd values of the expansion order $q$ (here $\beta J=5$ and $\Gamma/J=\pm 1/2$). Bottom: Severity of the sign problem as measured by the weighted sign $\langle \text{sgn} \rangle$ as a function of $\beta J$ on a log-linear scale for different values of the coupling parameter $\Gamma/J$. For negative $\Gamma$ values $\langle \text{sgn} \rangle=1$, whereas for positive values the quantity decays exponentially fast. (The lines are to guide the eye.)  } 
\end{figure}

For models with no sign problem and hence strictly non-negative weights, we have $\langle \text{sgn}\rangle=1$. Conversely, a severe sign problem corresponds to $\langle \text{sgn}\rangle \approx 0$, which stems from approximately equal amounts of negative and positive weights. Figure~\ref{fig:sign} illustrates this in the context of the spin triplet model: the top panel shows the average weights for different orders of $q$ for both the sign-problematic ($\Gamma>0$) and the sign-problem-free ($\Gamma<0$) cases for one set of parameters $\{\beta, \Gamma, J\}$. The bottom panel depicts the behavior of the figure of merit $\langle \text{sgn}\rangle$ as a function of $\beta J$ for various values of $\Gamma/J$. As is evident from the figure, the sign problem becomes more and more severe as the system gets colder, decaying exponentially with inverse temperature $\beta$. For the evaluation of thermal averages, which is carried out via Eq.(\ref{eq:A}), the decay of $\langle \text{sgn}\rangle$ implies exponentially slow convergence rates due to $\langle \text{sgn} \rangle$ appearing in the denominator, which results in highly fluctuating quantities with diverging error bars.

Before moving on, it should be made clear that the appearance of the sign problem is not an artifact of the flavor of the quantum Monte Carlo algorithm being used to sample the configuration space---in this case, ODE. Rather, the sign problem would similarly appear in any other standard QMC algorithm. Nonetheless, we will argue in the following section that the sign problem is more amenable to treatment within the framework of ODE. There, the sign problem has a very clear signature: negative-valued weights only appear when $\Gamma>0$ and for odd expansion orders. This observation will be exploited to devise a mechanism for grouping (or, re-summing) ODE weights in a way that completely eliminates negative-valued weights. 

\subsection{Resolution of the sign problem}
We now consider the resolution of the sign problem for our toy model by addressing the following question. Is there a decomposition of the partition function into easily computable and consistently positive weights? 
We answer this question in the affirmative by devising a method for grouping together ODE configurations in a specific manner to form what we refer to as `grouped configurations.' These grouped configurations will in turn produce grouped weights, which, as we shall see, are strictly positive. We note that the idea of grouping together QMC weights in order to resolve or mitigate the sign problem is of course not new and has been applied with varying degrees of success to other physical models in the context of other QMC algorithms (see, e.g., Refs.~\cite{resum1,resum2}). 

Since we expect the sign problem to be maximally severe in the low-temperature ($\beta \to \infty$) limit, we will determine the grouping based on the weights in that limit. We will then show that solving the sign problem in this limit 
also solves the sign problem in all other settings. At low temperatures, the weight Eq.~(\ref{eq:Wm0m1}) simplifies to (see App.~\ref{app:wc}):
\beq\label{eq:Wm0m1b}
W^{(\LT)}_{\{E_0^{\otimes m_0},E_1^{\otimes m_1}\}}= \frac{ (-\Gamma)^q \beta^{m_0-1}}{\Delta^{m_1}(m_0-1)!} \e^{-\beta E_0} \,.\nonumber\\
\eeq
We note in passing that the existence of a low-temperature weight is a non-trivial issue. It originates from ODE not being an expansion in the inverse temperature, $\beta$ but rather in the off-diagonal parameter $\Gamma$. As such, the various terms in the expansion are complete functions of $\beta$; hence, the series can be successfully used at arbitrary temperatures. Nonetheless, the weight, Eq.~(\ref{eq:Wm0m1b}), still changes sign with the parity of $q=m_0+m_1-1$. 
The above property of the low-temperature weight suggests the grouping together of weights with a \emph{fixed} number of excited states, (that is, a fixed value of $m_1$) and an increasing number of ground states ($m_0$) \emph{ad infinitum}. 

For the sake of simplicity, we shall construct grouped configurations by assembling randomly chosen configurations from each $(m_0,m_1)$ sector (i.e., with $m_0$ ground-state configurations and $m_1$ excited-state configurations).  
For any given $m_1$, a grouped configuration ${\mathcal{C}_{(m_1)}}$ will thus be a sequence of standard ODE configurations with an increasing number of ground states. Importantly, the smallest $m_0$ for which there exists a configuration with a given $m_1$ is \hbox{$m_0^{(\text{init})}=m_1-1$} (with the exception of $m_1=0$ for which \hbox{$m_0^{(\text{init})}=1$}). This immediately implies that the smallest order $q^{(\text{init})}=m_0^{(\text{init})}+m_1-1$ of any given grouped configuration is even, corresponding to a positive-valued initial standard weight. 

The weight of a grouped configuration $\mathcal{C}_{(m_1)}$ is thus simply
$
 W_{\mathcal{C}_{(m_1)}}=W_{(m_1)}/N_{(m_1)}
 $
 where
 \beq\label{eq:Wm1}
 W_{(m_1)}=\sum_{m_0}^{\infty} N_{(m_0,m_1)} W_{(m_0,m_1)}
 \eeq
 is the total weight of all configurations with $m_1$, and $N_{(m_1)}= \prod_{m0}^{\infty}N_{(m_0,m_1)}$ is the number of grouped configurations within that sector. We note here that since standard weights decay combinatorially fast with $m_0$, the evaluation of $W_{(m_1)}$ requires in practice summing only a finite number of terms.
 
To check whether the weight of a grouped configuration is positive,  we first observe that the sign of the grouped weight $W_{\mathcal{C}_{(m_1)}}$ is the sign of $W_{(m_1)}$ which is in itself a sum of standard weights with alternating signs. To show that this sum is always positive, we evaluate it in 
the worst-case scenario, i.e., the low-temperature limit, where it can be analytically computed to be:
 \bea
 W_{(m_1)}^{(\LT)}&=&\sum_{m_0} N_{(m_0,m_1)} W^{(\LT)}(m_0,m_1)\\\nonumber
&=&\frac{(3\beta \Gamma^2)^{m_1-1}\left[ 3\beta^2 \Gamma^2 + m_1(m_1-1)\right] }{\e^{\beta(E_0+2\Gamma)}\beta \Delta^{m_1}m_1!}\,.
\eea
We have therefore shown that a grouped ODE weight is a strictly positive quantity. 

Along with the positivity of the grouped configuration, it is equally important to show that the re-summed weights are efficiently computable. The sum in Eq.~(\ref{eq:Wm1}) is in principle an infinite one. However, the ODE weights decay approximately as $(\beta \Gamma)^q/q!$ where $q$ is the expansion order (see Ref.~\cite{ODE}). The weight decay is also evident in Fig.~\ref{fig:sign}(top): The distribution of weights is centered around $\langle q\rangle \propto \beta \Gamma$ with a width that is on the order of $\sigma_q \propto \sqrt{\beta \Gamma}$. To obtain $W_{(m_1)}$ it is therefore enough to sum over $O(\sqrt{\beta \Gamma})$ terms around $\langle q\rangle$. 

\subsection{QMC simulations: \\grouped vs. standard ODE}
We are now in a position to compare the performance of a QMC algorithm sampling the grouped ODE weights introduced above against those of standard ODE QMC. To do that, we importance-sample the respective configuration spaces of the two algorithms  for equal amounts of computation time. To do that, we importance-sample standard ODE $10^5$ times, and sample grouped-ODE for the same duration (which corresponds to approximately $10^5/\sqrt{\beta \Gamma}$ grouped-ODE samples). 

We examine the thermal average of the diagonal energy \hbox{$H_c/J=Z_1 Z_2 + Z_2 Z_3 +Z_3 Z_1$} for different values of $\beta J$ and $\Gamma/J$. The evaluation of $\langle H_c/J\rangle$, which is a diagonal operator, is done simply by assigning every ODE configuration $\mathcal{C}=\{|z\rangle, S_q\}$ an associated value, namely $E_z/J$ (where $E_z$ is the classical energy of $|z\rangle$). The evaluation of $\langle H_c/J\rangle$ for a grouped configuration similarly follows from the grouped configurations being weighted sums of standard ODE configurations.
For grouped ODE, we importance-sample the $m_1$ sectors with probabilities proportional to $W_{(m_1)}$, Eq.~(\ref{eq:Wm1}), and then randomly choose grouped configurations from the chosen sector with equal probabilities. 
For standard ODE, we sample standard configurations with probabilities proportional to the absolute value of $W_{(E_0^{\otimes m_0},E_1^{\otimes m_1})}$ given in Eq.~(\ref{eq:Wm0m1}). 
Since in the standard QMC case we sample the configuration space according to the wrong (absolute-valued)  distribution, we invoke Eq.~(\ref{eq:A}), to obtain correct thermal averages. 

\begin{figure*}
\subfigure[]{\includegraphics[width=0.95\columnwidth]{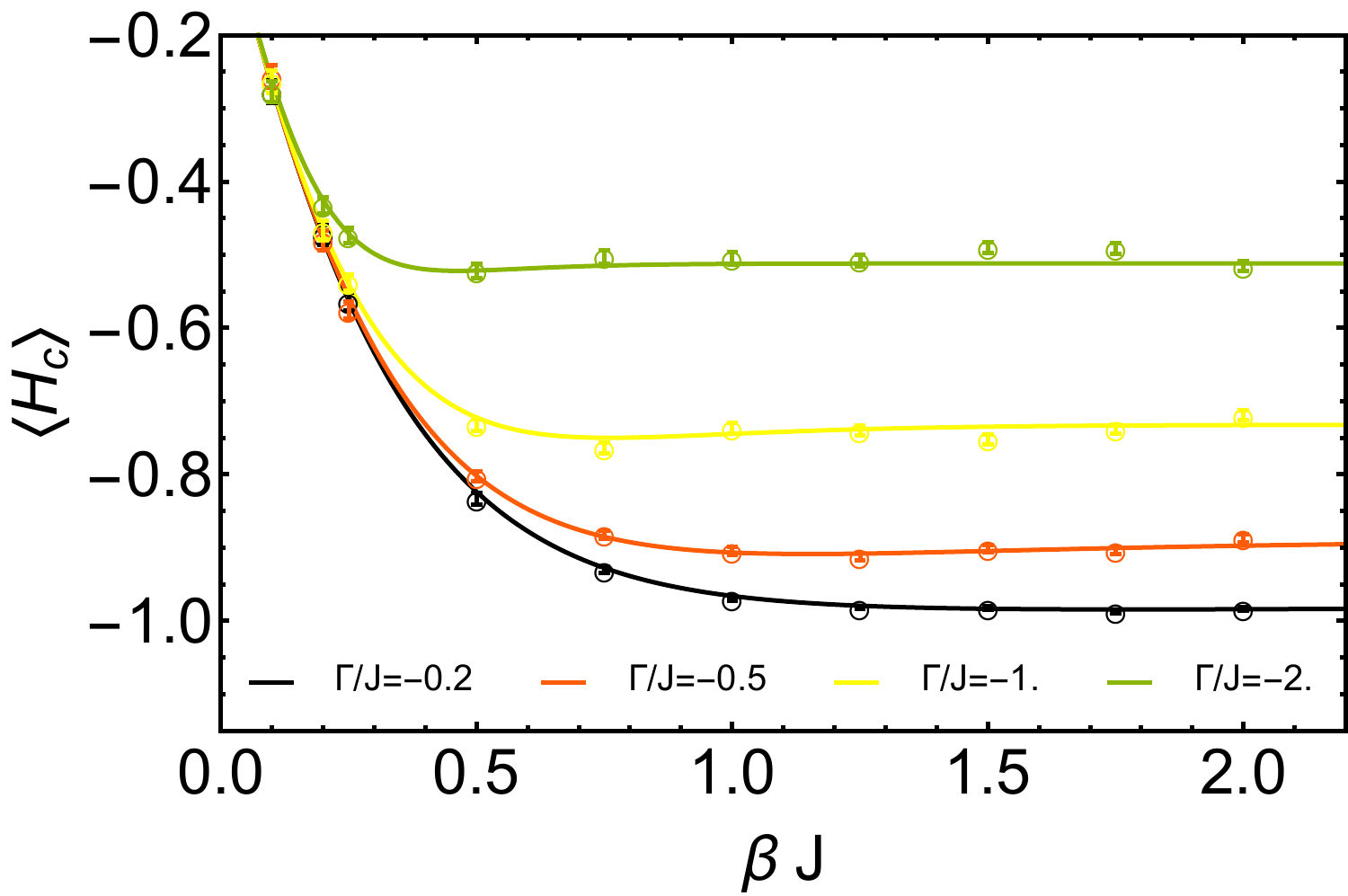}}
\subfigure[]{\includegraphics[width=0.95\columnwidth]{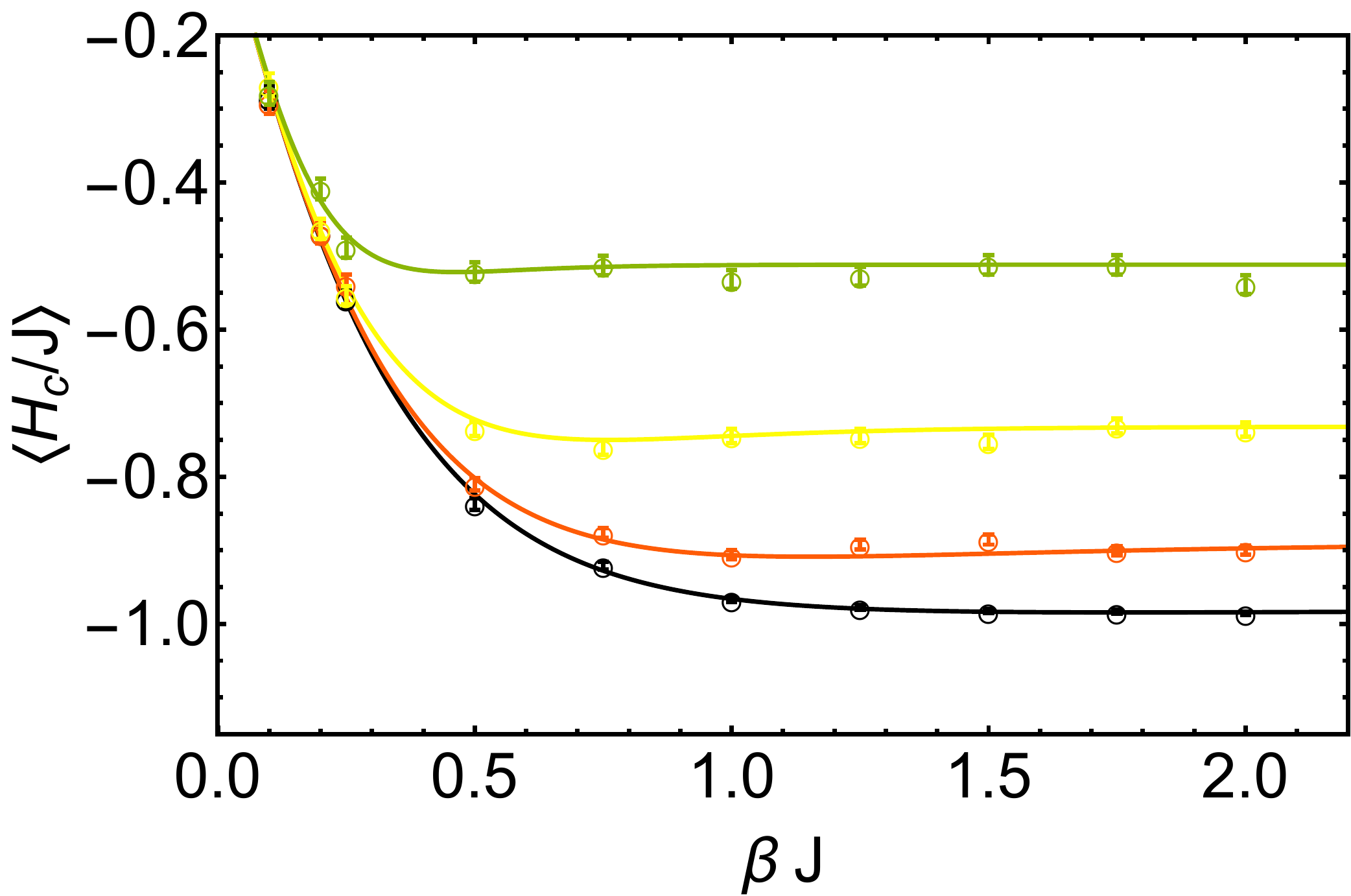}}
\subfigure[]{\includegraphics[width=0.95\columnwidth]{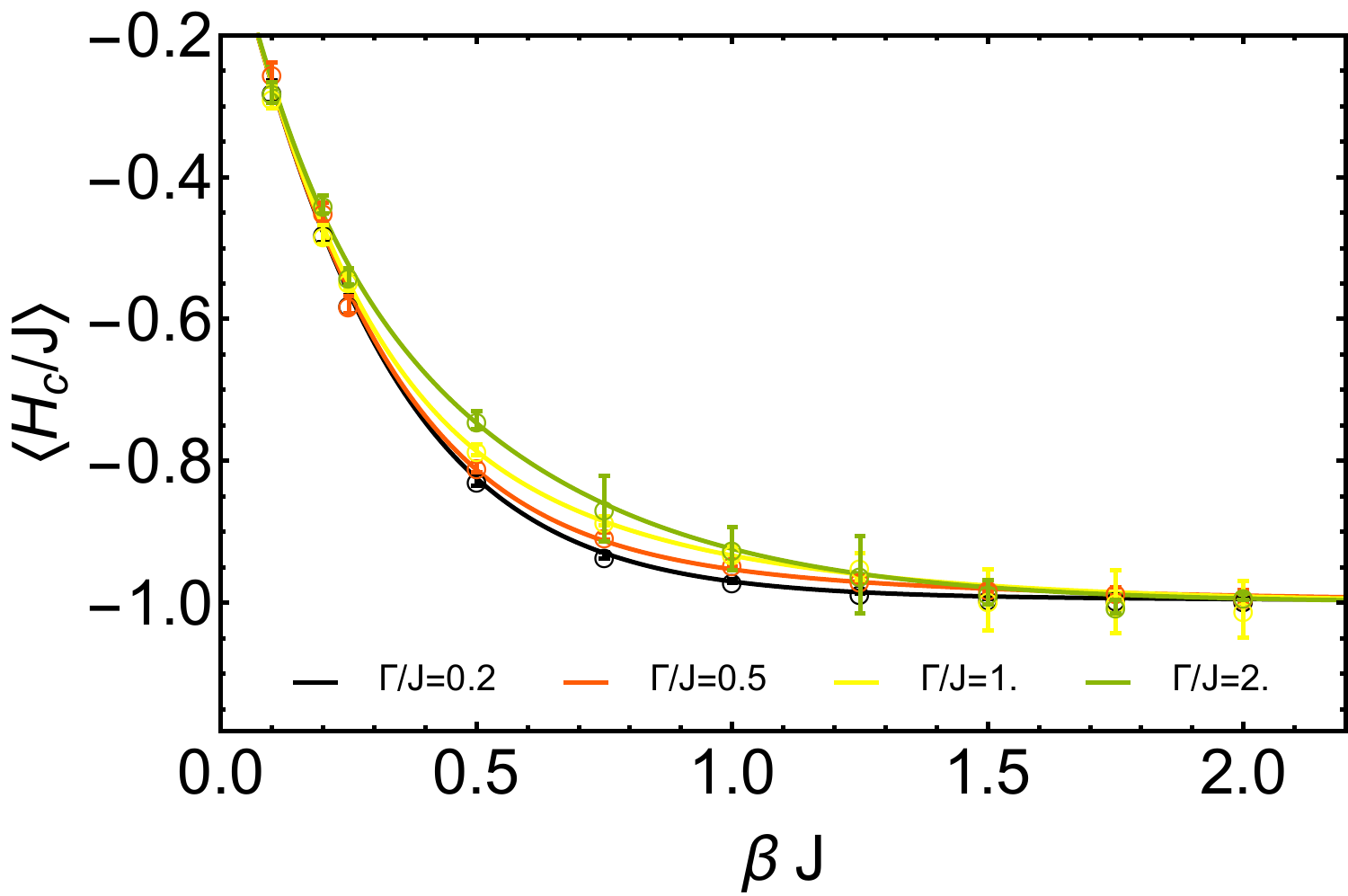}}
\subfigure[]{\includegraphics[width=0.95\columnwidth]{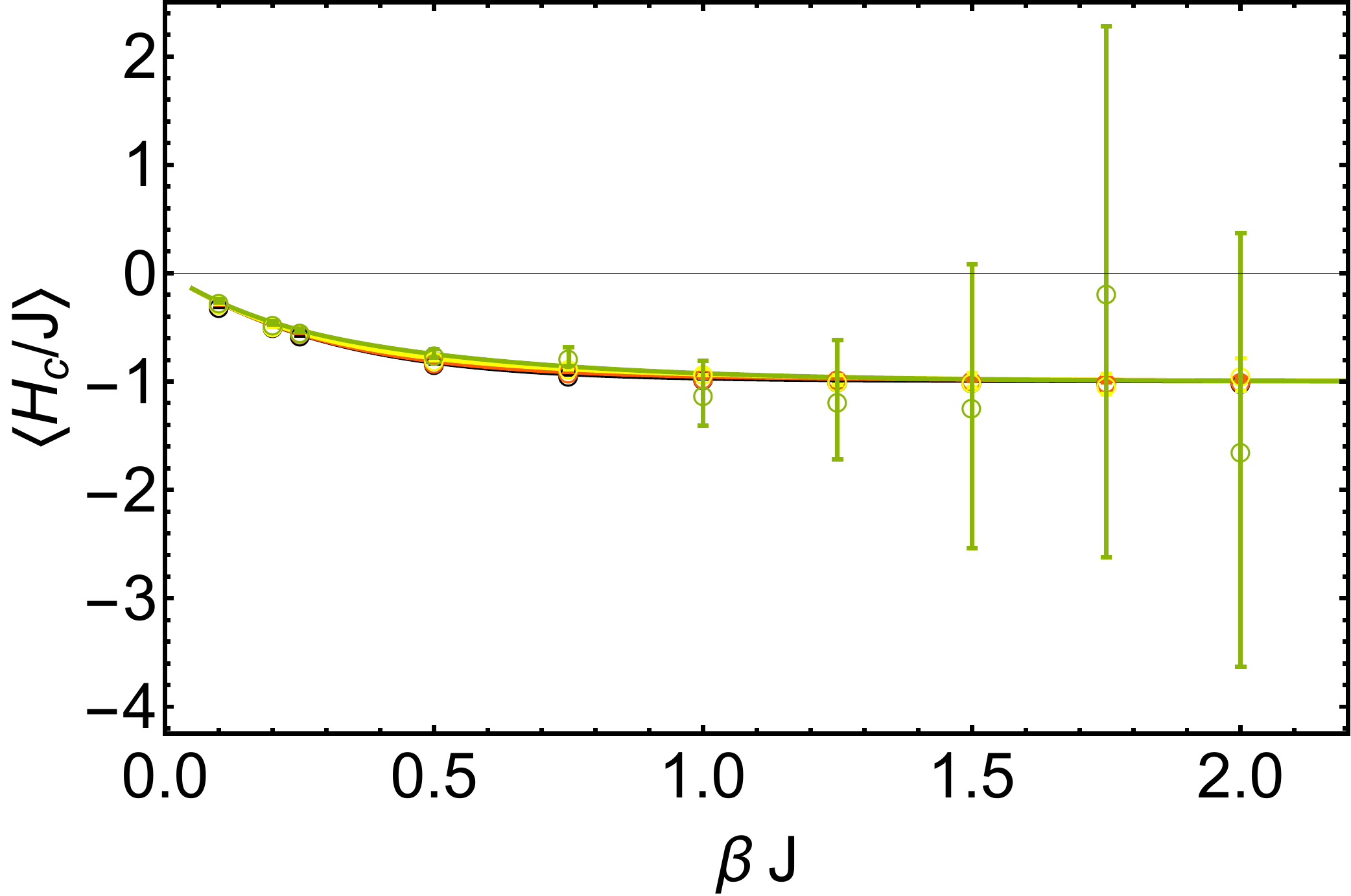}}
\caption{\label{fig:res} QMC thermal averages of the diagonal energy $\langle H_c/J \rangle $  as a function of $\beta J$ for different values of $\Gamma/J$. The solid lines are exact-diagonalization results. (a) Grouped ODE averages for negative values of $\Gamma$ (no sign problem). (b) Standard ODE for negative values of $\Gamma$ (no sign problem). (c) Grouped ODE averages for positive values of $\Gamma$ (sign problem). (d) Standard ODE for positive values of $\Gamma$ (sign problem). While for $\Gamma<0$ 
[(a)-(b)] both standard ODE and grouped ODE perform similarly well, the performances of the two algorithms differ considerably in the sign problematic case [(c)-(d)]. While grouped ODE encounters no problems, in contrast, standard ODE diverges (note the difference in the vertical scales between the two cases).}
\end{figure*}

The results of the simulations are summarized in 
Fig.~\ref{fig:res}. The two top panels, Fig.~\ref{fig:res}(a)-(b) show $\langle H_c/J \rangle$ as a function of $\beta J$ for several $\Gamma<0$ values. Here, the model is sign-problem-free and both algorithms perform similarly well. The two bottom panels, Fig.~\ref{fig:res}(c)-(d) depict the performance of the two algorithms in the presence of a sign problem, i.e., for positive $\Gamma$. Now, standard ODE weights oscillate---and more rapidly so with increasing values of $\beta J$. This leads to to diverging $\langle H_c/J\rangle$ averages and correspondingly, very large error bars [Fig.~\ref{fig:res}(d)]. In contrast, the grouped ODE algorithm does not encounter that problem, leading to efficient sampling, and in turn, a decent evaluation of the thermal average [Fig.~\ref{fig:res}(c)].
  
\section{Summary and discussion}
In the preceding section, we presented a mechanism for generating positive-valued weights for the quantum Monte Carlo simulation of a sign-problematic frustrated triplet of spin-$1/2$ particles. We have shown that a systematic regrouping of off-diagonal expansion~\cite{ODE} weights allows for the efficient importance-sampling of configuration space, thereby resolving the sign problem for that model. 
The single most-important remaining open question is whether or not the method can be extended to apply to large-scale spin systems for which exact diagonalization techniques are no longer feasible. 

We address this question by re-examining the technique in the context of large many-body systems. We start by noting that a standard ODE expansion can readily be carried out for large-scale systems, leading to a sign problem for positive $\Gamma$ values, similar to the frustrated spin triplet case. 
For general models however, the spectrum of the classical component of the Hamiltonian $H_c$ will consist of  multiple energy levels $E_0<E_1<E_2<\ldots$ and the standard ODE weight will be a function of their multiplicities $m_0, m_1, m_2,\ldots$ (to be compared with the two-level spectrum of the spin triplet).

As demonstrated above, to resolve the sign problem, it suffices to consider the behavior of the weights in the low-temperature limit where the sign problem is most pronounced. 
Interestingly, the low-temperature weight in the general case is a straightforward generalization of the one given by Eq.~(\ref{eq:Wm0m1b}) for the spin-triplet case, namely,
\beq\label{eq:Wm0m1c}
W^{(\LT)}_{\{E_0^{\otimes m_0},E_1^{\otimes m_1},\ldots\}}= \frac{ (-\Gamma)^q \beta^{m_0-1}}{(\prod_{j\neq0} \Delta^{m_j}_j) (m_0-1)!} \e^{-\beta E_0} \,,\nonumber\\
\eeq
where $m_0$ denotes the multiplicity of the lowest energy appearing in the path, $E_0$,  the expansion order is $q$ given by \hbox{$q=\sum_j m_j -1$} and $\Delta_j=E_j-E_0$ is the gap to the $j$th energy level (see App.~\ref{app:wc}).
Positive-valued grouped ODE weights can thus be constructed by collecting together standard weights with fixed excited-state multiplicities $m_j$ (for all $j>0$)  and an increasing number of minimal energy configurations $m_0$. Similar to the spin triplet model, we can expect the number of configurations within any given 
$\{ m_j \} $ sector to grow (to leading order) exponentially with $m_0$; i.e., \hbox{$N_{(m_0,m_1,m_2,\ldots)} \approx \alpha(m_1,m_2,\ldots)^{m_0}$} for some positive $\alpha(m_1,m_2,\ldots)$. This allows us to evaluate (at least approximately) the weight of a grouped ODE configuration, explicitly, 
\bea
&W&^{(\LT)}_{(m_1,m_2,\ldots)} = \sum_{m_0=m_{+}-1}^{\infty} N_{(m_0,m_1,m_2,\ldots)}  W^{(\LT)}_{(m_0,m_1,m_2,\ldots)} \nonumber\\
&\approx&\alpha \e^{-\beta(E_0+\alpha \Gamma)} \frac{(-\Gamma)^{m_+}}{\prod_j \Delta_j^{m_j}}\frac{\gamma(m_+-2,-\alpha \beta \Gamma)}{(m_+-1)!} \,,
\eea
where we have defined $m_{+}\equiv \sum_{j>0} m_j$, and $\gamma(\cdot,\cdot)$ is the lower incomplete Gamma function. 
The above quantity is strictly positive in every $\{m_{j}\}$ sector, as was the case for our toy model in the preceding section. We may therefore expect grouped ODE weights to be positive for general sign-problematic spin models as well. 

It should be noted nonetheless, that the existence of positive-valued weights is not the only factor in a QMC simulation and, depending on the specifics of the model being studied, update steps based on the properties of the grouped ODE configurations may need to be devised for the simulation to take place.

\section{Conclusions}

The sign problem is one of the most fundamental bottlenecks of quantum Monte Carlo simulations of many-body physics, chemistry and material sciences~\cite{Wiese-PRL-05,marvianHenLidar,signProbSandvik}. Any progress made towards its resolution is therefore of importance to the general scientific community. 

In this study we presented a technique for resolving the sign problem for spin models, based on the regrouping of weights from the off-diagonal expansion QMC algorithm. 
Using an easily diagonalizable toy model, we show that in the QMC simulation of a bonafide sign-problematic system it is indeed possible to efficiently carry out the simulation 
if the standard off-diagonal expansion QMC weights are grouped together in a particular manner, based on the properties of the weights in the low-temperature limit. 
We also presented general arguments as to our technique's prospects for a successful extension to larger spin systems as well. We leave that for future research. 

\begin{acknowledgements}
The research is based upon work (partially) supported by the Office of
the Director of National Intelligence (ODNI), Intelligence Advanced
Research Projects Activity (IARPA), via the U.S. Army Research Office
contract W911NF-17-C-0050. The views and conclusions contained herein are
those of the authors and should not be interpreted as necessarily
representing the official policies or endorsements, either expressed or
implied, of the ODNI, IARPA, or the U.S. Government. The U.S. Government
is authorized to reproduce and distribute reprints for Governmental
purposes notwithstanding any copyright annotation thereon.
\end{acknowledgements}

\bibliography{refs}
\appendix

\begin{widetext}
\section{\label{app:wc}Derivation of ODE weights}

We provide below a brief summary of the concept of divided differences which is a recursive division process. This method is typically encountered when calculating the coefficients in the interpolation polynomial in the Newton form.
The divided differences~\cite{dd:67,deboor:05} of a function $f(\cdot)$ with $(q+1)$ distinct inputs $[x_0,\ldots,x_q]$ is defined as
\beq\label{eq:divideddifference2}
f[x_0,\ldots,x_q] \equiv \sum_{j=0}^{q} \frac{f(x_j)}{\prod_{k \neq j}(x_j-x_k)}\,.
\eeq
The above expression is well-defined even if the inputs have repeated values, in which case one must resort to a limiting process. 
A divided difference can alternatively be defined via the following recursion relations which also provide.a simple way to evaluate it. 
\beq\label{eq:ddr2}
f[x_i,\ldots,x_{i+j}] = \frac{f[x_{i+1},\ldots , x_{i+j}] - f[x_i,\ldots , x_{i+j-1}]}{x_{i+j}-x_i} \,,
\eeq
with $i\in\{0,\ldots,q-j\},\ j\in\{1,\ldots,q\}$ with the initial conditions
$f[x_i] = f(x_{i})$ with $i \in \{ 0,\ldots,q \}$.
In the case where there are multiplicities, i.e., repeated values, of the input values and assuming without loss of generality that $x_0<x_1<x_2<\ldots$, we find~\cite{dd:67,deboor:05}:
\beq
f[\{x_0,m_0\},\ldots,\{x_r,m_r\}]=f[\underbrace{x_0,\ldots,x_0}_{(m_0+1) \, \text{times}},\ldots,\underbrace{x_r,\ldots,x_r}_{(m_r+1) \, \text{times}}]=\frac1{\prod_{j=0}^{r} m_j!}\prod_{j=0}^{r} \frac{\partial^{m_j}}{\partial x_j^{m_j}} f[x_0,\ldots,x_r]
\eeq
where $(m_j+1)$ is the multiplicity of the value $x_j$ ($m_j \geq 0$ for all $j=0\ldots r$). 

For the purposes of this study, we focus on the exponential function:
\beq
f[x_0,\ldots,x_q]=e^{-\beta[x_0,\ldots,x_q]} 
\eeq
which appears in the expression for the ODE weight, Eq.~(\ref{eq:gbw}). 
Here, 
\bea
f[x_0,\ldots,x_r]=\e^{-\beta[x_0,\ldots,x_r]}=\sum_{i=0}^r \frac{\e^{-\beta x_i}}{\prod_{j\neq i} (x_i-x_j)}
\eea
is the function with inputs without the multiplicities ($m_j=0$). 
With multiplicities, we have:
\beq
f[\{x_0,m_0\},\ldots,\{x_r,m_r\}]=\sum_i \frac1{\prod_{j=0}^{r} m_j!}\left( \prod_{j=0}^{r} \frac{\partial^{m_j}}{\partial x_j^{m_j}} \frac{\e^{-\beta x_i}}{\prod_{j\neq i} (x_i-x_j)}\right)
\eeq
We note that
\beq
\frac{\partial^{m_j}}{\partial x_j^{m_j}} \e^{-\beta x_j} = (-\beta)^{m_j} \e^{-\beta x_j}
\eeq
and for any integer $a>0$
\beq
\frac{\partial^{k}}{\partial x_i^{k}}\frac1{(x_i-x_j)^a} = (-1)^{k}\frac{(a+k-1)!}{(a-1)!(x_i-x_j)^{a+k}} \,.
\eeq
for $i \neq j$. In particular for $a=1$:
\beq
\frac{\partial^{m_j}}{\partial x_j^{m_j}}\frac1{x_i-x_j} = \frac{m_j!}{(x_i-x_j)^{m_j+1}} \,.
\eeq 
Differentiating the $i$th term of $f[\cdot]$ above with respect to all $x_j$ for $j \neq i$, we get:  
\beq
f[\{x_0,m_0\},\ldots,\{x_r,m_r\}]=\sum_i \frac1{\prod_{j=0}^{r} m_k!}\left(  \frac{\partial^{m_i}}{\partial x_i^{m_i}} \frac{m_j!\e^{-\beta x_i}}{\prod_{j\neq i} (x_i-x_j)^{m_j+1}}\right) \,,
\eeq
which simplifies to
\bea
f[\{x_0,m_0\},\ldots,\{x_r,m_r\}]=\sum_i \frac1{m_i!}\left(  \frac{\partial^{m_i}}{\partial x_i^{m_i}} \frac{\e^{-\beta x_i}}{\prod_{j\neq i} (x_i-x_j)^{m_j+1}}\right)\,.
\eea
The derivative with respect to $x_i$ for the $i$th term gives, using the chain rule
\beq
f[\{x_0,m_0\},\ldots,\{x_r,m_r\}]=\sum_i \frac1{m_i!}
\left(  \sum_{k_i=0}^{m_i} {m_i \choose k_i} \frac{\partial^{m_i-k_i}\e^{-\beta x_i}}{\partial x_i^{m_i-k_i}} \frac{\partial^{k_i}}{\partial x_i^{k_i}} \frac1{\prod_{j\neq i} (x_i-x_j)^{m_j+1}} \right)
\eeq
which simplifies to
\beq
f[\{x_0,m_0\},\ldots,\{x_r,m_r\}]=\sum_i \e^{-\beta x_i} \frac{1}{m_i!}\left(  \sum_{k_i=0}^{m_i}  \frac{m_i!}{(m_i-k_i)! k_i!} (-\beta)^{m_i-k_i} 
\sum_{\substack{k_{j\neq i}=0 \\ \text{s.t.} \,\sum_{j\neq i} k_j=k_i}}^{k_i} k_i!\prod_{j\neq i}  \frac{(-1)^{k_j}}{k_j!} \frac{(m_j+k_j)!}{m_j!(x_i-x_j)^{m_j+k_j+1}}\right)\,.\nonumber
\eeq
Further simplifications give
\beq
f[\{x_0,m_0\},\ldots,\{x_r,m_r\}]=\sum_i \e^{-\beta x_i} (-\beta)^{m_i} \left(  \sum_{k_i=0}^{m_i} \frac{\beta^{-k_i} }{(m_i-k_i)!} \sum_{\substack{k_{j\neq i}=0 \\ \text{s.t.} \,\sum_{j\neq i} k_j=k_i}}^{k_i} \prod_{j\neq i} \frac1{(x_i-x_j)^{m_j+k_j+1}}{m_j+k_j \choose m_j} \nonumber
\right)\,.
\eeq
Making the substitution $k_i \to m_i-k_i$, we arrive at
\beq
f[\{x_0,m_0\},\ldots,\{x_r,m_r\}]=\sum_i \e^{-\beta x_i} (-1)^{m_i} \left(  \sum_{k_i=0}^{m_i} \frac{\beta^{k_i} }{k_i!} \sum_{\substack{k_{j\neq i}=0 \\ \text{s.t.} \,\sum_{j\neq i} k_j=m_i-k_i}} \prod_{j\neq i} \frac1{(x_i-x_j)^{m_j+k_j+1}}{m_j+k_j \choose m_j} 
\right)\,. \nonumber
\eeq
Denoting $x_i-x_j=\Delta_{ij}$ we get:
\beq \label{eq:f}
f[\{x_0,m_0\},\ldots,\{x_r,m_r\}]=\sum_i \e^{-\beta x_i} (-1)^{m_i} \left(  \sum_{\sum_j k_j=m_i} \frac{\beta^{k_i} }{k_i!} \prod_{j\neq i} \frac1{\Delta_{ij}^{m_j+k_j+1}}{m_j+k_j \choose m_j} 
\right)\,.
\eeq

\subsection{Case of only two repeated values}

In the two energy-level case, where only $m_0$ and $m_1$ appear the expression below simplifies to
\bea
f[\{x_0,m_0\},\{x_1,m_1\}]&=&\e^{-\beta x_0} (-1)^{m_0} \left(  \sum_{k_0=0}^{m_0} \frac{\beta^{k_0} }{k_0! \Delta^{m_1+m_0-k_0+1}}{m_1+m_0-k_0 \choose m_1} \right)\\
&+&\e^{-\beta x_1} (-1)^{m_1} \left(  \sum_{k_1=0}^{m_1} \frac{\beta^{k_1} }{k_1! (-\Delta)^{m_0+m_1-k_1+1}}{m_0+m_1-k_1 \choose m_0} \right)\nonumber\,,
\eea
where we have denoted $\Delta=\Delta_{10}=x_1-x_0$. Further simplification of the sums above and denoting $q=m_0+m_1-1$ we get
\beq
f[\{x_0,m_0\},\{x_1,m_1\}]=\e^{-\beta x_0} \frac{(m_0+m_1)!\left[{}_1F_1(1-m_0,1-q,-\beta \Delta)-\e^{-\beta \Delta} {}_1F_1(1-m_1,1-q,\beta \Delta )\right]}{(-1)^{m_1}\Delta^q(m_0-1)!(m_1-1)!} \,,\nonumber
\eea
where  $ {}_1F_1$ is the Kummer confluent hypergeometric function, as asserted in the main text, Eq.~(\ref{eq:Wm0m1}).

\subsection{Low temperature limit}
In the low-temperature limit, where $\beta \to \infty$, only the term proportional to $\e^{-\beta x_0} \beta^{m_0}$ in Eq.~(\ref{eq:Wm0m1}) survives giving 
\beq\label{eq:Wm0m1cApp}
f^{(\LT)}\{x_0,m_0\},\ldots,\{x_r,m_r\}]= \frac{ (-1)^q \beta^{m_0-1}}{(\prod_{j\neq0} \Delta^{m_j}_{0j}) (m_0-1)!} \e^{-\beta E_0} \,,
\eeq
as asserted in the main text,  Eq.~(\ref{eq:Wm0m1c}). Equation~(\ref{eq:Wm0m1b}) is a particular case of the above equation in the case of only two energy levels. 

\section{\label{app:Nm0m1}Derivation of $N_{(m_0,m_1)}$}
Here we derive the expression for the number of distinct configurations (within a given parity sector) having the energy multiplicities $(m_0,m_1)$
\beq\label{eq:Mm0m1app}
N_{(m_0,m_1)}=- 2(-1)^{m_0} \delta_{0,m_1}
+ \frac{3^{m_1} 2^{m_0-m_1}(m_0-1)!}{6 (m_0-m_1+1)! m_1!}
 \left[
4 m_1(m_1-1)+3(m_0-m_1+1)(m_0-m_1)  \right]
\eeq
given in the main text.

As discussed in the main text, ODE configurations can be described as closed paths on the hypercube of classical states. In the spin-triplet case, the paths consist of moves between four points of a given parity. The even parity states are {\bf 0, 3, 5} and {\bf 6}, the last three of which are ground states contributing to the $m_0$ count and the fourth to $m_1$ count. Similarly, {\bf 1, 2, 4} and {\bf 7} are the odd parity states, the first three of which contributing to $m_0$ and the last to $m_1$.

Starting with the simpler case of $m_1=0$, the number of distinct configurations with $m_0$ ground states corresponds to enumerating the number of sequences of {\bf 3}, {\bf 5} and {\bf 6} (alternatively {\bf 1, 2} and {\bf 4}) of length $m_0$ obeying the constraints that the first and last state in each sequence are the same and no two adjacent states can be the same. This gives:
\beq
N_{(m_0,0)} = 2 \left( 2^{m_0 - 2} - (-1)^{m_0}\right) \,.
\eeq 
Next, we consider cases with a nonzero number of $m_1$ excited states (i.e., {\bf 0} states, if one restricts to the even parity sector). Since {\bf 0} states must be separated by at least one ground state, we can identify two types of sequences. Sequences of the first type begin and end with a {\bf 0} state. Thhey must have the form ${\bf 0 * 0  \ldots 0 * 0}$, where {\bf*} denotes a sequence of ground states.
For any given $m_1$, the number of ground-state sequences between any two {\bf 0} states is $K=m_1-1$. The lengths of these sequences, $k_i$, with $i=1\ldots K$ must sum to $m_0$. The number of possible ground-state sequences of length $k_i$ is $3 \times 2^{k_i-1}$. For this case, we can thus write
 \beq
N^{(\text{I})}_{(m_0,m_1)} = \sum_{\sum_{k_i}=m_0} {m_0\choose k_1 \ldots k_{m_1-1}} \times 3 \cdot 2^{k_i-1} =
{m_0-1 \choose m_1 -2} \times 3^{m_1-1} \cdot 2^{m_0-m_1+1}\,.
\eeq 
The second type of sequences has the general form ${\bf * 0 * 0 \ldots 0 * 0 *}$. Here, the number of ground-state sequences is $K=m_1+1$ augmented with the constraint that the first state of the first sequence and the last state of the last sequence must be the same. As in the other case, 
the number of sequences of $k_i$ consecutive ground states is $3 \times 2^{k_i-1}$ except for the last sequence for which there are only $2^{k_{m_1+1}-1}$ due to the additional constraint. We thus obtain:
 \beq
N^{(\text{II})}_{(m_0,m_1)} = \frac1{3} \sum_{\sum_{k_i}=m_0} {m_0\choose k_1 \ldots k_{m_1+1}} \times 3 \times 2^{k_i-1} = {m_0-1 \choose m_1} \times 3^{m_1+1} \cdot 2^{m_0-m_1+1} \,.
\eeq 
All three expressions above may be combined to a single expression, Eq.~(\ref{eq:Mm0m1app}) above. 

\end{widetext}
\end{document}